\documentclass[10pt, reprint, twocolumn, nofootinbib, showkeys, aps, prd, showpacs, superscriptaddress]{revtex4-2}
\usepackage{amsmath}
\usepackage{amssymb}
\usepackage[export]{adjustbox}
\usepackage{fix-cm}
\usepackage{adjustbox}
\usepackage{latexsym,amsmath,amssymb}
\RequirePackage{graphicx}
\RequirePackage{mathptmx}
\usepackage[arrowdel]{physics}
\usepackage{graphicx,epstopdf}
\usepackage{slashed}
\usepackage{bm}
\usepackage{cases}
\usepackage{color}
\usepackage{bm}
\usepackage{dcolumn}
\usepackage{lipsum}
\usepackage{diagbox}
\usepackage{array}
\usepackage{multirow}
\usepackage{makecell}
\usepackage[colorlinks = true, linkcolor = blue, urlcolor  = blue, citecolor = red, anchorcolor = blue]{hyperref}
\begin{document}
\title{On the symmetries of the modified Emden-type equation}
\author{Subhra Mondal}
%\altaffiliation[Also at ]{Physics Department, XYZ University.}%Lines break automatically or can be forced with \\
 \email{cosmology313@gmail.com}
\author{Amitava Choudhuri}
 \email{amitava\_ch26@yahoo.com}
\affiliation{Department of Physics, The University of Burdwan, Golapbag, Purba Bardhaman, West Bengal 713104, India}
%\pacs{02.30.Hq, 02.30.IK, 05.45.-a}
\begin{abstract}
For an autonomous system, the Lagrangian symmetries are embedded in the symmetries of the differential equation. Recently, it has been found that modified Emden-type equations follow non-standard Lagrangian functions that involve neither the kinetic energy term nor the potential function. By working with one such Lagrangian, we have calculated the Lagrangian symmetries and explicitly demonstrated that, as in the case of standard Lagrangian functions, the variational symmetries of the non-standard Lagrangian are also included in the Lie symmetries of the non-linear differential equation. The Lie algebra is also studied. The symmetry-based solutions to the equation are also derived using invariant curve conditions.
\end{abstract}
\maketitle
\tableofcontents
\section{Introduction}\label{Intro}
A widely studied model in nonlinear dynamics is the \textit{modified Emden-type equation}, often referred to as the \textit{modified Painlevé-Ince equation} \cite{ince1956ordinary} is given by
\begin{equation}\label{modEMDEN}
    \ddot x + \alpha x \dot x + \beta x^3=0 \,\,,\hspace{1cm} x \equiv x(t)\,\,.
\end{equation} Here and throughout the article a dot over a variable like $x$ denotes a single differentiation with respect to time $t$ and so on. The parameters $\alpha$ and $\beta$ are taken to be arbitrary. The frictional coefficient $\alpha$ is positive and satisfies the inequality $0<\alpha<1$. This dissipative-like equation received serious attention from both mathematicians and physicists. For more than a century, it appeared in many mathematical problems including the realization of uni-valued functions as defined by second-order differential equations \cite{golubev1950lectures} and Riccati equation \cite{chisholm1987class}, on the other hand, it plays an important role in several applicative contexts \cite{chandrasekhar1957introduction,chandrasekar2007general,almeida1992lie,leach1985first,dixon1990solutions,rogers1984wave,mcvittie1933mass,mcvittie1967gravitational,mcvittie1984elliptic,yang1954conservation,choudhuri2015spatially,nandi2024symmetry}. In solving the non-linear ordinary differential Eq. (\ref{modEMDEN}), several advancements have been made in different methods analytically \cite{chandrasekar2007general,ghosh2013factorization,biswas2019analysis,chandrasekar2005unusual,mohanasubha2018symmetries,senthilvelan2015symmetries} and numerically \cite{mukherjee2011solution} for any values of $\alpha$ and $\beta$ over the past decade. The equation in (\ref{modEMDEN}) has a two-parameter Lie point symmetry group for any values of $\alpha$ and $\beta$. However, when selecting $\beta=\frac{\alpha^2}{9}$, it exhibits eight-parameter Lie point symmetries \cite{mahomed1985linear,mahomed1989lie,chandrasekar2007general,choudhuri2015spatially}. In this case, we write the modified Emden-type equation as
\begin{equation}\label{modEMDENnew}
    \ddot x + 3 k x \dot x + k^2 x^3=0\,\,.
\end{equation}    
It is straightforward to verify that Eq. (\ref{modEMDENnew}) can be linearized by using the coordinate transformation
\begin{equation}\label{coord_trans}
     y(t)=e^{k\int x(t)dt}\,\,,
\end{equation}
and also that it follows from a Lagrangian \cite{carinena2005lagrangian}
\begin{equation}\label{lagEMDEN}
     \mathcal{L}=\frac{1}{\dot x+ k x^2}\,\,.
\end{equation}     
For any autonomous differential equation as that in Eq. (\ref{modEMDEN}) or (\ref{modEMDENnew}) the standard Lagrangian is defined by $\mathcal{L}=\mathcal{T}-\mathcal{V}$, where $\mathcal{T}$ is the kinetic energy of the system modeled by the equation and $\mathcal{V}$, the corresponding potential function. The Lagrangian in (\ref{lagEMDEN}) is not of this form. Thus it is often called the non-standard Lagrangian \cite{musielak2008standard}.
\par So far as the standard Lagrangians are concerned, there exists plenty of literature on the computation of Lagrangian symmetries of physical systems using the so-called Noether's theorem which states that the symmetries of a variational problem yield conservation laws. The Lie point symmetry of a differential equation is a set of transformations of the dependent and independent variables that leave the equation unchanged \cite{olver1993applications}. The number of variational symmetries is less than the number of Lie symmetries of an equation. In fact, the Lagrangian symmetries form a subset of the set of Lie symmetries of the differential equation.
\par Basically, the two approaches rely on the idea of invariance concerning an infinitesimal transformation of the dynamical variables. In the case of Noether's theorem, the entity that remains invariant is the Action Integral, while for the Lie method, it pertains to the equations of motion. The latter approach is less limiting compared to the former and offers a larger array of invariants and/or accommodates a broader range of problems to be addressed. To illustrate the broader applicability of the Lie method, it is seen in a study \cite{leach1981applications} for a one-dimensional linear system, when restricting to point transformations, Noether's theorem yields five symmetry generators, while the Lie method generates eight. Nevertheless, a more significant limitation of Noether's theorem arises when examining multi-dimensional systems. It fails to produce the Jauch-Hill-Fradkin (J-H-F) tensor for the harmonic oscillator, nor does it provide the Laplace-Runge-Lenz (L-R-L) vector for the classical Kepler problem. In contrast, the Lie method does offer these results \cite{prince1980symmetries,prince1981lie}.
\par In this work we shall employ Noether's theorem \cite{Noether1918} to study the variational symmetries of Eq. (\ref{modEMDENnew}) for the non-standard Lagrangian in (\ref{lagEMDEN}) and demonstrate that these symmetries are also embedded in the set of corresponding Lie symmetries. The approach we follow has an old root in classical mechanics literature. For example, in 1951 Hill \cite{hill1951hamilton} provided a simplified account of Noether's theorem by considering the infinitesimal transformations of the dependent and independent variables characterizing a classical field. In classical mechanics, the variational symmetry of a system having   degrees of freedom is a consequence of the invariance of the action functional
\begin{equation}\label{action}
   \mathcal A=\int_{t_1}^{t_2}\mathcal{L}(\vec{x},\, \dot{\vec{x}},\, t)\,dt 
\end{equation}
under the infinitesimal transformations of independent and dependent variables
\begin{equation}
    t'= t+\delta t\,\,\,\,{\rm  and}\,\,\,\,x'=x+\delta x\,\,,
\end{equation}
such that
\begin{equation}\label{inf_trans}
    \delta t=\epsilon \xi(\vec{x},t)\,\,\,\,{\rm  and}\,\,\,\,\delta x_i=\epsilon \eta_i(\vec{x},t)\,\,,
\end{equation}
retaining only lowest-order terms in $\epsilon$ ($\epsilon<<1$), and the infinitesimals $\xi(\vec{x},t)$
and $\eta_i(\vec{x},t)$ are the arbitrary functions
that give the generator of the infinitesimal transformations. These transformations are generated by the vector field
\begin{equation}\label{vector_field}
    X=\xi(\vec{x},\,t)\frac{\partial}{\partial t}+\sum_{i=1}^n\eta_i(\vec{ x},\,t)\frac{\partial}{\partial x_i}\,\,.
\end{equation}
Here $\vec{x}=\{x_1,\,x_2,\,....\,x_n\}$ and $\dot{\vec{x}}=\frac{d\vec{x}}{dt}=\{\dot{x_1},\,\dot{x_2},\,....\,\dot{x_n}\}$ represent the generalized coordinates and velocities of the system. The prolonged infinitesimal generator of order $m$ corresponding to $X$ in Eq. (\ref{vector_field}) is written as \cite{olver1993applications}
\begin{equation}
    X^{(m)}=X+X_1+X_2+...+X_m\,\,,
\end{equation}
with
\begin{subequations}\label{prolong}
    \begin{align}
        X^{(1)} & = X+X_1=X+\sum_{i=1}^n \eta_i^{(1)}\frac{\partial}{\partial \dot x_i}\,\,, \label{1st_prolong}\\
        X^{(2)} & = X+X_1+X_2=X^{(1)}+\sum_{i=1}^n\eta_i^{(2)}\frac{\partial}{\partial \ddot{\vec{x}}}\label{2nd_prolong}
  \end{align}
\end{subequations}
and so on. Clearly, $X^{(1)}$  and $X^{(2)}$ are the first and second prolongations of $X$. In writing (\ref{prolong}) we used
\begin{subequations}
    \begin{align}
    \eta_i^{(1)} & =\left[\dot\eta_i(\vec{x},t)- \dot\xi(\vec{x},t)\dot{\vec{x}}\right] \label{eta'}\\
    \eta_i^{(2)} &=\left[\ddot\eta_i(\vec{x},t)-2\dot\xi(\vec{x},t)\ddot x_i- \ddot\xi(\vec{x},t)\dot x_i\right]\,\,.\label{eta"}
\end{align}
\end{subequations}
The action functional remains invariant under those point transformations whose constituents  $\xi$ and $\eta_i$ satisfy \cite{choudhuri2008symmetries}
\begin{equation}\label{I_dot}
    \frac{d\mathcal{I}}{dt}+\sum_{i=1}^{n}\left(\xi\dot x_i-\eta_i\right)\left(\frac{\partial \mathcal{L}}{\partial x_i}-\frac{d}{dt}\frac{\partial \mathcal{L}}{\partial \dot x_i}\right)=0
\end{equation}
with $\mathcal{I}$ given by
\begin{equation}\label{I}
    \mathcal{I}=\sum_{i=1}^{n}\left(\xi\dot x_i-\eta_i\right)\frac{\partial \mathcal{L}}{\partial \dot x_i}-\xi \mathcal{L}+f(\vec x,t)\,\,,
\end{equation}
where the gauge function $f(\vec{x},\,t)$ satisfies the differential equation
\begin{equation}\label{f_dot}
    \frac{df(\vec{x},\,t)}{dt}=\xi \mathcal{L}+ \xi \frac{\partial \mathcal{L}}{\partial t}+\sum_{i=1}^n\left(\eta_i\frac{\partial \mathcal{L}}{\partial t}+(\dot\eta_i- \dot\xi\dot x_i)\frac{\partial}{\partial \dot x_i}\right)\,\,.
\end{equation}
Along the trajectories of the system, the Euler-Lagrange equations hold good such that the second term in Eq. (\ref{I_dot}) is zero. Therefore
\begin{equation}\label{I_conserved}
     \frac{d\mathcal{I}}{dt}=0.
\end{equation}
This implies $\mathcal{I}$ given in (\ref{I}) is a conserved quantity. This invariant quantity together with the differential equation satisfied by the gauge function is commonly stated as the Noether's theorem. 
\par In Sect. \ref{Noether_sym} we work out the variational symmetries of the modified Emden-type equation by using nonstandard Lagrangian (\ref{lagEMDEN})  and present results for the corresponding conserved quantities. In Sect. \ref{Lie_sym} we present results for  Lie symmetries of the equation and demonstrate how the variational symmetries are embedded in the symmetries of the differential equation. We also find all possible Lie symmetry-based solutions to the differential equation. Finally, we make concluding remarks in Sect. \ref{con_rem}.
%%%%%%%%%%%%%%%%%%%%%%%%%%%%%%%%%%%%%%%%%%%%%%%%%%%%%%%%%%%%%%%%%%%%%%%%%%%%%%%%%%%%%%%%%%%
\section{Noether symmetry analysis}\label{Noether_sym}
To study the variational or Noether's symmetries of (\ref{modEMDENnew}), we begin with (\ref{I}) which for our (1+1) dimensional system is
\begin{equation}\label{Inew}
    \mathcal{I}=\left(\xi\dot x-\eta\right)\frac{\partial \mathcal{L}}{\partial \dot x}-\xi \mathcal{L}+f(x,t)\,\,.
\end{equation}
We substitute the expression of Lagrangian from (\ref{lagEMDEN}) in (\ref{Inew}) and take the time derivative of the resulting expression of conserved quantity to zero (see Eq. \ref{I_conserved}) reads
\begin{eqnarray}\label{res_exp}
    2 k x \eta+ p^3 f_x+2 k p^2f_x+k^2 p x^4f_x+ p\eta_x-\nonumber \\ 2p^2\xi_x- k p x^2\xi_x+ p^2f_t+2 k p x^2f_t +\nonumber \\ k^2x^4f_t+\eta_t-2p\xi_t-k x^2\xi_t = 0\,\,.
\end{eqnarray}
Here and afterward the suffices on $f$, $\eta$, and $\xi$ denote partial derivatives with respect to the appropriate variables and $p=\dot x$. In deriving (\ref{res_exp}) we have made use of (\ref{modEMDENnew}). Equation (\ref{res_exp}) can be globally satisfied for any particular choice of $p$ provided the sum of $p$ independent terms, the coefficients of linear, quadratic, and cubic terms in $p$ vanish separately. This viewpoint leads to four determining equations for the gauge function $f$ and infinitesimal generators $\xi$ and $\eta$ of Noether symmetry. In particular, we have
\begin{subequations}\label{det_eqnNOETHER}
\begin{align}
 2 k x \eta+ k^2x^4f_t+\eta_t-k x^2\xi_t & =0 \,\,,\\
 k^2 x^4f_x+\eta_x-k x^2\xi_x+2 k x^2f_t -2\xi_t & = 0 \,\,,\\
2 k f_x-2\xi_x +f_t & =0 \,\,,\\
f_x & =0\,\,.
\end{align}
\end{subequations}
Solving equations in (\ref{det_eqnNOETHER}), we find the following gauge function and infinitesimal Noether symmetry generator
\begin{subequations}\label{sol_of_det_eqnNOETHER}
\begin{align}
f(x,t) & = a(t)=\frac{1}{4}c_0k^2t^4-c_2k t^3+c_3+c_4t+c_5t^2\,\,,\\
\xi(x,\,t) & = \frac{c_0}{2}(k^2 t^3 x-k t^2)+c_1+c_2(t-\frac{3}{2}k t^2 x)+ \nonumber\\
    &\phantom{={}} \frac{1}{2}c_4 x+c_5 t x\,\,,\\
\eta(x,t) & = c_0(1-2k t x+\frac{3}{2}k^2 t^2 x^2-\frac{1}{2}k^3 t^3 x^3)+ \nonumber\\
    &\phantom{={}} c_2(2 x-3 k t x^2+\frac{3}{2} k^2 t^2 x^3)-\frac{c_4}{2} k x^3+ \nonumber\\
    &\phantom{={}} c_5 (x^2-k t x^3)\,\,,
\end{align}
\end{subequations}
respectively, where $c_i$ are arbitrary integration constants. We, therefore, have a five-parameter set of solutions from which we can construct five linearly independent group generators given by
\begin{subequations}\label{generators_NOETHER}
\begin{align}
X_1 & =(k^2 t^3 x-k t^2)\frac{\partial}{\partial t}+(2-4k t x+3k^2 t^2 x^2-k^3 t^3 x^3)\frac{\partial}{\partial x}\,\,,\\
X_2 & =\frac{\partial}{\partial t}\,\,,\label{time_tran}\\
X_3 & =(2t-3k t^2 x)\frac{\partial}{\partial t}+(4 x-6 k t x^2+3 k^2 t^2 x^3)\frac{\partial}{\partial x}\,\,,\\
X_4 & = x \frac{\partial}{\partial t}- k x^3\frac{\partial}{\partial x}\,\,,\\
X_5 & =t x\frac{\partial}{\partial t}+(x^2-k t x^3)\frac{\partial}{\partial x}\,\,.
\end{align}
\end{subequations}
These five one-parameter operators generate a five-parameter $\mathfrak{sl}(2,\mathbb{R})\bigoplus _s 2A_1$ Lie algebra $\Lambda^5$ and satisfy the closure property. The commutation relations between these symmetry generators are given in \autoref{tab:Table1}. The Lie algebra $\Lambda^5$ depicted in \autoref{tab:Table1} contains the following 14 subalgebras: $(X_1), (X_2), (X_3), (X_4), (X_5), (X_1, X_3), (X_2, X_4),(X_3, X_4), (X_3, X_5),\\ (X_4, X_5), (X_1, X_3, X_5), (X_2, X_4, X_5),(X_3, X_4, X_5), (X_2, X_3, X_4, X_5)$.\\ The first integrals corresponding to generators in (\ref{generators_NOETHER}) can be found from (\ref{Inew}), (\ref{sol_of_det_eqnNOETHER}), and (\ref{generators_NOETHER}) are given by
\begin{subequations}\label{first_int}
\begin{align}
I_1 & = \frac{(1-k t x)(1+k t(pt-x)+k^2t^2x^2)}{(p+k x^2)^2}+a(t)\,\,,\\
I_2 & = -\frac{(2p+kx^2)}{(p+k x^2)^2}+a(t)\,\,,\label{JacobiInt}\\
I_3 & = \frac{p(-2t+3kt^2x)+x(2-4ktx+3k^2t^2x^2)}{(p+kx^2)}+a(t)\,\,,\\
I_4 & = -\frac{x}{(p+kx^2)}+a(t)\,\,,\\
I_5 & = \frac{x^2-2p t x-2k t x^3}{(p+k x^2)^2}+a(t)\,\,.
\end{align}
\end{subequations}
\begin{widetext}
{\small\begin{center}
\renewcommand{\arraystretch}{1.5}
\begin{table}[ht]
\caption{Commutation table for the Noether symmetry generators. Each element $X_{ij}$ in the Table is represented by $X_{ij}=[X_i,\,X_j]$.}
\begin{tabular}{|c|c|c|c|c|c|}
\hline 
\diagbox[width=.9cm, height=.9cm, innerleftsep=1.5mm, innerrightsep=1.5mm]{$X_i$}{$X_j$}
&$X_1$&$X_2$ &$X_3$&$X_4$&$X_5$ \\
\hline 
$X_1$&$0$&$k X_3$ &$4 X_1$&$2 (X_2-kX_5)$&$X_3$\\
\hline
$X_2$&$-k X_3$&$0$ &$2 (X_2-3kX_5)$&$0$&$X_4$\\
\hline
$X_3$&$-4 X_1$&$-2 (X_2-3kX_5)$&$0$&$2 X_4$&$4 X_5$\\
\hline
$X_4$&$-2 (X_2-kX_5)$&$0$&$-2 X_4$&$0$&$0$\\
\hline
$X_5$&$-X_3$&$-X_4$ &$-4 X_5$&$0$&$0$\\
\hline
\end{tabular}
\label{tab:Table1}
\end{table}
\end{center}}
\end{widetext}
The Lagrangian of the system is explicitly time-independent and, as expected, (\ref{JacobiInt}) represents the energy function or Jacobi integral of the modified Emden-type equation given in (\ref{modEMDENnew}) and the corresponding symmetry generator is the time translation operator (\ref{time_tran}). It is worth mentioning that $\frac{\partial}{\partial t}$ does not necessarily imply energy integral in all cases, c.f. angular momentum, L-R-L vector and J-H-F tensor (for a detailed discussion see \cite{leach1981applications}).
%%%%%%%%%%%%%%%%%%%%%%%%%%%%%%%%%%%%%%%%%%%%%%%%%%%%%%%%%%%%%%%%%%%%%%%%%%%%%%%%%%%%%%%%
\section{Lie symmetry analysis}\label{Lie_sym}
Here we are interested in calculating the symmetries of the modified Emden-type equation (\ref{modEMDENnew}) which follows from non-standard inverse Lagrangian (\ref{lagEMDEN}) and examine how variational symmetries calculated in Sect. \ref{Noether_sym} are embedded in the symmetries of the differential equation. A second-order $(1+1)$ dimensional ordinary differential equation of the form $\Psi(t,\,x,\,\dot x,\,\ddot x)=0$ is invariant under the twice extended group with infinitesimals $(\xi ,\eta, \eta^{(1)}, \eta^{(2)})$, and possesses Lie point symmetries provided the invariance condition \cite{olver1993applications,cantwell2002introduction,Bairagi_Symmetry}
\begin{equation}\label{diff_eqn}
    X^{(2)}\Psi(t,\,x,\,\dot x,\,\ddot x)=0\,\,,
\end{equation}
where $X^{(2)}$ is the second-order prolongation of the vector field $X$ obtained by specializing vector field (\ref{vector_field}) to $(1+1)$ degrees of freedom. In our case
\begin{equation}
    \Psi(t,\,x,\,\dot x,\,\ddot x)=\ddot x + 3 k x \dot x + k^2 x^3\,\,.
\end{equation}
Now, in order to find the determining equations, we use $(1+1)$ dimensional forms of (\ref{inf_trans}), (\ref{1st_prolong}), (\ref{2nd_prolong}), (\ref{eta'}) and (\ref{eta"}) in the invariance condition (\ref{diff_eqn}) and also equating the coefficients of linear, quadratic, and cubic terms in $p$ of the resulting expression to zero separately in a similar manner showed in the previous Sect. \ref{Noether_sym}. Finally, the obtained determining equations are
\begin{subequations}\label{det_eqn_LIE}
\begin{align}
\xi_{xx} & = 0\,\,, \\
6kx\xi_x + \eta_{xx}-2\xi_{tx} & = 0\,\,, \\
3k\eta+3k^2x^3\xi_x+3kx\xi_t+2\eta_{tx}-\xi_{tt} & = 0\,\,, \\
3k^2x^2\eta-k^2x^3\eta_x+3kx\eta_t+2k^2x^3\xi_t+\eta_{tt} & = 0\,\,.
\end{align}
\end{subequations}
By solving equations in (\ref{det_eqn_LIE}), we find the infinitesimal generators of Lie symmetry
\begin{subequations}\label{inf_gen_LIE}
\begin{align}
        \xi(x,t) & = d_1 t + d_2 + d_3 k t^2 x + d_4 x + d_5 tx + d_6 (2t^2-k t^3 x) + \nonumber\\
    &\phantom{={}} d_7 (3k t^2-k^2 t^3x) + d_8 (-2kt^3+k^2t^4x)\,\,,\\
        \eta(x,t) & = -d_1 x + d_3 (-2 x+2 k t x^2-k^2 t^2 x^3)  - d_4 k x^3 + \nonumber\\
    &\phantom{={}} d_5 (x^2-k t x^3) + d_6 (2 t x-3k t^2 x^2+k^2t^3x^3) +\nonumber\\
    &\phantom{={}} d_7 (2-3k^2t^2x^2+k^3t^3x^3) + \nonumber\\
    &\phantom{={}} d_8 (4t-6k t^2x+4k^2t^3x^2-k^3t^4x^3) \,\,.
\end{align}
\end{subequations}
where $d_i$ are arbitrary integration constants. The infinitesimal generators (\ref{inf_gen_LIE}) have an eight-parameter set of solutions that lead to eight linearly independent Lie point symmetries
\begin{subequations}\label{Symmetry_generators_LIE}
\begin{align}
X_{L1} & = t\frac{\partial}{\partial t}-x\frac{\partial}{\partial x}\,\,, \\
X_{L2} & = \frac{\partial}{\partial t}\,\,, \\
X_{L3} & = k t^2 x\frac{\partial}{\partial t}+(-2 x+2 k t x^2-k^2 t^2 x^3)\frac{\partial}{\partial x}\,\,, \\
X_{L4} & = x\frac{\partial}{\partial t}-k x^3\frac{\partial}{\partial x}\,\,, \\
X_{L5} & = t x\frac{\partial}{\partial t}+(x^2-k t x^3)\frac{\partial}{\partial x}\,\,, \\
X_{L6} & = (2t^2-k t^3 x)\frac{\partial}{\partial t}+(2 t x-3k t^2 x^2+k^2t^3x^3)\frac{\partial}{\partial x}\,\,, \\
X_{L7} & = (3k t^2-k^2 t^3x)\frac{\partial}{\partial t}+(2-3k^2t^2x^2+k^3t^3x^3)\frac{\partial}{\partial x}\,\,, \\
X_{L8} & = (-2kt^3+k^2t^4x)\frac{\partial}{\partial t}+(4t-6k t^2x+4k^2t^3x^2-k^3t^4x^3)\frac{\partial}{\partial x}\,\,.
\end{align}
\end{subequations}
These eight one-parameter generators construct an eight-parameter Lie algebra $\Lambda^8$ obeying the closure property. The subscript $L$ on $X$  has been used merely to indicate that the vector fields in (\ref{Symmetry_generators_LIE}) are Lie symmetry generators of (\ref{modEMDENnew}). The differential equation being analyzed possesses the maximal number of eight Lie point symmetries, indicating that the associated Lie symmetry algebra is $\mathfrak{sl}(3,\mathbb{R})$, which can be clearly confirmed through the vector field generators presented in (\ref{Symmetry_generators_LIE}) \cite{mahomed1990symmetry}. The commutation relations between these symmetry generators are given in \autoref{tab:Table2}. The Lie algebra $\Lambda^8$ depicted in \autoref{tab:Table2} contains 61 subalgebras as follows: $(X_{L1}), (X_{L2}), (X_{L3}), (X_{L4}), (X_{L5}), (X_{L6}), (X_{L7}), (X_{L8}),\\ (X_{L1}, X_{L2}), (X_{L1}, X_{L3}), (X_{L1}, X_{L4}), (X_{L1}, X_{L5}), (X_{L1}, X_{L6}),\\ 
(X_{L1}, X_{L7}), (X_{L1}, X_{L8}), (X_{L2}, X_{L4}), (X_{L3}, X_{L4}), (X_{L3}, X_{L5}),\\
(X_{L3}, X_{L6}), (X_{L3}, X_{L8}), (X_{L4}, X_{L5}),(X_{L5}, X_{L6}), (X_{L6}, X_{L8}),\\ 
(X_{L7}, X_{L8}), (X_{L1}, X_{L2}, X_{L4}), (X_{L1}, X_{L3}, X_{L4}), (X_{L1}, X_{L3}, X_{L5}),\\
(X_{L1}, X_{L3}, X_{L6}), (X_{L1}, X_{L3}, X_{L8}), (X_{L1}, X_{L4}, X_{L5}),
(X_{L1}, X_{L4}, X_{L8}),\\ (X_{L1}, X_{L5}, X_{L6}), (X_{L1}, X_{L6}, X_{L8}), 
(X_{L1}, X_{L7}, X_{L8}), (X_{L2}, X_{L4}, X_{L5}),\\ (X_{L3}, X_{L4}, X_{L5}),
(X_{L3}, X_{L5}, X_{L6}), (X_{L3}, X_{L6}, X_{L8}), (X_{L4}, X_{L5}, X_{L6}),\\ 
(X_{L5}, X_{L6}, X_{L8}), (X_{L6}, X_{L7}, X_{L8}),(X_{L1}, X_{L2}, X_{L4}, X_{L5}),\\ (X_{L1}, X_{L3}, X_{L4}, X_{L5}), (X_{L1}, X_{L3}, X_{L4}, X_{L8}), (X_{L1}, X_{L3}, X_{L5}, X_{L6}),\\
(X_{L1}, X_{L3}, X_{L6}, X_{L8}), (X_{L1}, X_{L4}, X_{L5}, X_{L6}), 
(X_{L1}, X_{L5}, X_{L6}, X_{L8}),\\ (X_{L1}, X_{L6}, X_{L7}, X_{L8}),
(X_{L2}, X_{L3}, X_{L4}, X_{L5}), (X_{L3}, X_{L4}, X_{L5}, X_{L6}),\\ 
(X_{L3}, X_{L5}, X_{L6}, X_{L8}), (X_{L3}, X_{L6}, X_{L7}, X_{L8}),\\ (X_{L1}, X_{L2}, X_{L3}, X_{L4}, X_{L5}),(X_{L1}, X_{L3}, X_{L4}, X_{L5}, X_{L6}),\\ 
(X_{L1}, X_{L3}, X_{L5}, X_{L6}, X_{L8}), (X_{L1}, X_{L3}, X_{L6}, X_{L7}, X_{L8}),\\
(X_{L1}, X_{L4}, X_{L5}, X_{L6}, X_{L8}), (X_{L1}, X_{L2}, X_{L3}, X_{L4}, X_{L5}, X_{L6}),\\
(X_{L1}, X_{L3}, X_{L4}, X_{L5}, X_{L6}, X_{L8}), (X_{L1}, X_{L3}, X_{L5}, X_{L6}, X_{L7}, X_{L8})$.\\ Ideally, all Noether's symmetries should lie inside the Lie point symmetries. But looking at (\ref{generators_NOETHER}) and (\ref{Symmetry_generators_LIE}) we see that all the generators of the Lagrangian symmetry are not of the same form as those associated with the invariance of the equation. However, it is straightforward to verify that variational symmetries that do not appear in (\ref{generators_NOETHER}) can always be expressed as linear combinations of Lie symmetries (\ref{Symmetry_generators_LIE}). For example, we can write $X_1=X_{L7}-2k X_{L6}$ and $X_3=2X_{L1}-3X_{L3}$.
\begin{widetext}
{\small\begin{center}
\renewcommand{\arraystretch}{1.5}
\begin{table}[ht]
\caption{Commutation table for the Lie symmetry generators. Each element $X_{Lij}$ in the Table is represented by $X_{Lij}=[X_{Li},\,X_{Lj}]$.}
\begin{tabular}{|c|c|c|c|c|c|c|c|c|}
\hline 
\diagbox[width=1.2cm, height=1cm, innerleftsep=1.5mm, innerrightsep=1.5mm]{$X_{Li}$}{$X_{Lj}$}
&$X_{L1}$&$X_{L2}$ &$X_{L3}$&$X_{L4}$&$X_{L5}$&$X_{L6}$&$X_{L7}$&$X_{L8}$ \\
\hline 
$X_{L1}$&$0$&$- X_{L2}$ &$0$&$-2 X_{L4}$&$- X_{L5}$&$X_{L6}$&$X_{L7}$&$2 X_{L8}$\\
\hline
$X_{L2}$&$X_{L2}$&$0$ &$2k X_{L5}$&$0$&$ X_4$&$4 X_{L1}-3 X_{L3}$&$3k (2 X_{L1}- X_{L3})$&$2 (-3k X_{L6}+ X_{L7})$\\
\hline
$X_{L3}$&$0$&$-2k X_{L5}$&$0$&$-2 X_{L4}$&$-2 X_{L5}$&$0$&$2 (-2k X_{L6}+ X_{L7})$&$2 X_{L8}$\\
\hline
$X_{L4}$&$2 X_{L4}$&$0$&$2 X_{L4}$&$0$&$0$&$2 X_{L5}$&$2 (- X_{L2}+ 3k X_{L5})$&$-4 X_{L1}$\\
\hline
$X_{L5}$&$X_{L5}$&$- X_{L4}$ &$2 X_{L5}$&$0$&$0$&$0$&$-2 X_{L1}+ 3 X_{L3}$&$-2 X_{L6}$\\
\hline
$X_{L6}$&$- X_{L6}$&$-4 X_{L1}+3 X_{L3}$ &$0$&$2 X_{L5}$&$0$&$0$&$-X_{L8}$&$0$\\
\hline
$X_{L7}$&$- X_{L7}$&$-2 (-3k X_{L6}+ X_{L7})$&$-2 (-3k X_{L6}+ X_{L7})$&$-2 (- X_{L2}+ 3k X_{L5})$&$2 X_{L1}- 3 X_{L3} $&$X_{L8}$&$0$&$0$\\
\hline
$X_{L8}$&$- 2 X_{L8}$&$-2 (-3k X_{L6}+ X_{L7})$ &$- 2 X_{L8}$&$4 X_{L1}$&$2 X_{L6}$&$0$&$0$&$0$\\
\hline
\end{tabular}
\label{tab:Table2}
\end{table}
\end{center}}
\end{widetext}
\par To obtain group-invariant solutions of ordinary differential Eq. (\ref{modEMDENnew}) by exploiting Lie symmetry generators $X_{Li}$, we follow the \textit{invariant curve condition} approach \cite{hydon2000symmetry,cantwell2002introduction,Bairagi_Symmetry}. Let us consider any curve $\mathcal{C}$ on the $x$-$t$ plane and it is said to be invariant iff the tangent to curve $\mathcal{C}$ at each point is parallel to the tangent vector $(\xi(x,t), \eta(x,t))$. This invariant curve condition can be described mathematically through the introduction of a characteristic $ Q \overset{\text{def}}{=} \eta(x,t)- \dot x \xi(x,t)$. Every curve $\mathcal{C}$ on the $x$-$t$ plane that remains invariant under the group generated by $X_{Li}$ meets a crucial criterion called the invariant curve condition
\begin{eqnarray}\label{characteristic}
    \ Q \equiv \eta(x,t)- \dot x \xi(x,t)=0
\end{eqnarray}
on $\mathcal{C}$.
\begin{figure}[htbp!]
    \centering
    \includegraphics[width=86mm,height=65mm]{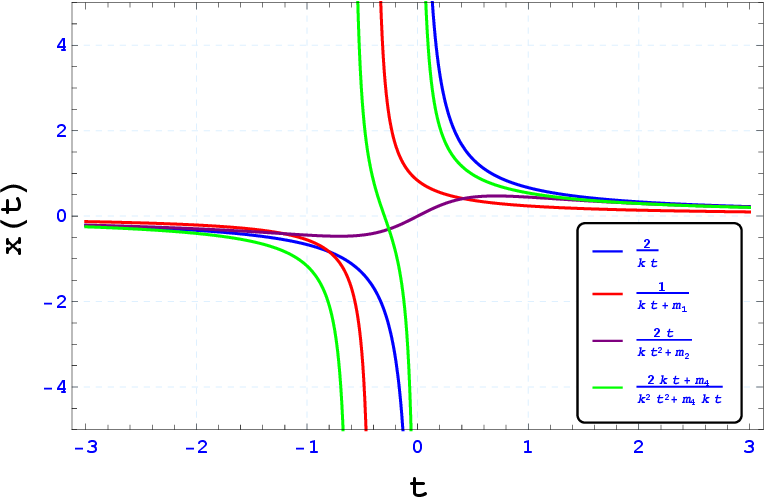}
    \caption{\small(Color online) Plot for Lie symmetry-based solutions to Eq. (\ref{modEMDENnew}) with typical values of constants $m_1=1.2$, $m_2=1.5$, $m_4=1.8$, and $k=3$}
    \label{solution plot emden}
\end{figure}
Now, we find non-trivial solutions for the modified Emden-type equation (\ref{modEMDENnew}) from the invariant curve condition (\ref{characteristic}) and each Lie symmetry generator in (\ref{Symmetry_generators_LIE}) which are listed as follows:
\begin{itemize}
    \item From $X_{L1}$: $x(t) = \frac{1}{kt}$ and $x(t) = \frac{2}{kt}$\,\,.

     \item From $X_{L4}$: $x(t) = \frac{1}{kt+m_1}$\,\,.

    \item From $X_{L5}$: $x(t) = \frac{2 t}{kt^2+m_2}$\,\,. 

     \item From $X_{L6}$: $x(t) = \frac{2}{kt}$ and $x(t) = \frac{2t}{kt^2+m_3}$\,\,.

     \item From $X_{L7}$: $x(t) = \frac{2}{kt}$\,\,.

     \item From $X_{L8}$: $x(t) = \frac{2}{kt}$ and $x(t) = \frac{2kt+m_4}{k^2t^2+m_4 kt}$\,\,.
\end{itemize}
Here $m_1, m_2, m_3, \text{and}\,\, m_4$ are arbitrary integration constants. Except for these other Lie symmetry generators, do not produce any non-trivial solution. We see that some of the solutions are identical, and obtain four unique solutions. The different solutions are plotted in Fig. \ref{solution plot emden}. In this context, we should note that in articles \cite{mohanasubha2018symmetries,senthilvelan2015symmetries}, a similar Lie symmetry-based solution of type $\frac{2kt+m_4}{k^2t^2+m_4 kt}$ has been presented. Some authors also produced another solution of type $\frac{1}{kt+m_1}$ applying different methodologies such as the factorization method \cite{ghosh2013factorization} and the power-law ansatz \cite{biswas2019analysis}.
%%%%%%%%%%%%%%%%%%%%%%%%%%%%%%%%%%%%%%%%%%%%%%%%%%%%%%%%%%%%%%%%%%%%%%%%%%%%%%%%%%
\section{Concluding remarks}\label{con_rem}
In this work, we calculated the variational symmetries of the modified Emden-type Eq. (\ref{modEMDENnew}) corresponding to the non-standard Lagrangian (\ref{lagEMDEN}) and explicitly demonstrated that as with symmetries of the standard Lagrangians,  the Noether's symmetries of (\ref{lagEMDEN}) are also embedded in the Lie symmetries of (\ref{modEMDENnew}). In this context, we note that (\ref{modEMDENnew}) also follows from another non-standard Lagrangian
\begin{equation}\label{newLAG}
    \mathcal{L'}=\sqrt{2 \dot x+k x^2}\,\,.
\end{equation}
It is of interest to note that the Lagrangians in (\ref{lagEMDEN}) and (\ref{newLAG}) are not connected by a gauge term. Yet, both yield the same equation of motion through the Euler-Lagrangian equation. Such Lagrangians are called the alternative Lagrangians. The presence of alternative Lagrangians in a physical system has deep consequences in physical theories \cite{morandi1990inverse, Talukdar2007}. For example, ambiguities can arise in associating symmetries with conservation laws. Moreover, the same physical system can lead to entirely different quantum mechanical systems via alternative Lagrangian descriptions. Thus it remains an interesting curiosity to compute Noether's symmetries using (\ref{newLAG}) and compare them with the results presented in this work.
%%%%%%%%%%%%%%%%%%%%%%%%%%%%%%%%%%%%%%%%%%%%%%%%%%%%%%%%%%%%%%%%%%%%%%%%%%%%%%%%%%%%%%%%%%%
\section*{Acknowledgments}
The authors express their gratitude to Prof. PGL Leach for his valuable suggestions. SM is thankful to the Government of West Bengal, India, for granting the State-funded Junior Research Fellowship (JRF).
%%%%%%%%%%%%%%%%%%%%%%%%%%%%%%%%%%%%%%%%%%%%%%%%%%%%%%%%%%%%%%%%%%%%%%%%%%%%%%%%%%%%%%%%%%%
\section*{Funding}
The authors wish to clarify that they have not received funding from institutional projects while preparing this article.
%%%%%%%%%%%%%%%%%%%%%%%%%%%%%%%%%%%%%%%%%%%%%%%%%%%%%%%%%%%%%%%%%%%%%%%%%%%%%%%%%%%%%%%%%%%%%%%%%
\section*{Data availability statement}
Data sharing does not apply to this article since no datasets were produced or examined during this study.
%%%%%%%%%%%%%%%%%%%%%%%%%%%%%%%%%%%%%%%%%%%%%%%%%%%%%%%%%%%%%%%%%%%%%%%%%%%%%%%%%%%%%%%%%%%
\section*{Code availability statement}
Code/Software sharing is not applicable to this article, as the current study did not involve generating or analyzing any code or software.
%%%%%%%%%%%%%%%%%%%%%%%%%%%%%%%%%%%%%%%%%%%%%%%%%%%%%%%%%%%%%%%%%%%%%%%%%%%%%%%%%%%%%%%%%%%
\section*{Declarations}
\subsection*{Conflict of interest}
The authors state that there are no conflicts of interest.
\newpage
\bibliography{reference}% Produces the bibliography via BibTeX.
\end{document}